\documentclass[%
 reprint,
 amsmath,amssymb,
 aps,
]{revtex4-2}
\usepackage{xcolor}%
\usepackage{graphicx}
\usepackage{dcolumn}
\usepackage{bm}

\newcommand{\pfr}[2]{\ensuremath{\frac{\partial #1}{\partial #2}}}
\newcommand{\pfi}[2]{\ensuremath{{\partial #1}/{\partial #2}}}

\newcommand{\Pra}{\textit{Pr}}
\newcommand{\mb}[1]{\mathbf{#1}}

\newcommand{\Ray}{Ra}

\newcommand{\eff}{\mathrm{eff}}

\newcommand{\ep}{\varepsilon}

\newcommand{\beq}{\begin{equation}}
\newcommand{\eeq}{\end{equation}}
\DeclareMathAlphabet\mathbfcal{OMS}{cmsy}{b}{n}

\begin{document}

\preprint{APS/123-QED}

\title{Hydrodynamic origin of Korteweg stresses from shear-induced horizontal buoyancy}

\author{Prabakaran Rajamanickam}
 \email{prabakaran.rajamanickam@manchester.ac.uk}
\affiliation{ Department of Mathematics, University of Manchester, Manchester M13 9PL, United Kingdom}%

\date{\today}

\begin{abstract}
A recent study~\cite{rajamanickam2025shear} of non-Boussinesq fluids in narrow channels identified a novel shear-induced horizontal buoyancy force that emerges upon depth-averaging the Navier--Stokes equations. This letter demonstrates that this force is formally equivalent to the divergence of a Korteweg stress tensor. Unlike classical Korteweg stresses, which are typically attributed to molecular-scale cohesive potentials or implemented through assumed constitutive relations, we show that this emergent stress arises purely from self-coupled transport where the internal Ostroumov flow is kinematically coupled to the local density gradient. We derive explicit expressions for the effective stress coefficients, revealing a fundamental dependence on the Prandtl number and Grashof number.  This correspondence is contrasted with classical Taylor dispersion, where the absence of self-coupling yields only a uniaxial stress. Although derived within a narrow-channel framework, our results establish a general hydrodynamic template for how quadratic gradient stresses can emerge from sub-scale, self-coupled flows, such as Marangoni or active-matter flows, offering a continuous transport-driven alternative to molecular mechanisms.
\end{abstract}

\maketitle


\section{Introduction}
\label{sec:intro}

When a horizontal density gradient is imposed in a narrow channel, the fluid cannot remain at rest~\cite{landau1987fluid}. The resulting differential hydrostatic pressure drives a steady circulation known as the Ostroumov flow, an internal circulation analogous to Hadley’s cell in atmospheric science. Characterised by antisymmetry about the channel’s mid-plane, this flow produces no net discharge in confined systems. The transport and mixing of the underlying scalar field, typically temperature or solute concentration, are profoundly influenced by this internal shear through a mechanism analogous to Taylor dispersion,
as was first identified in this context by Erdogan and Chatwin~\cite{erdogan1967effects}. Traditional investigations into these flows almost universally invoke the Boussinesq approximation, neglecting density variations in the inertial terms. However, recent work has shown that abandoning this approximation during depth-averaging reveals a novel phenomenon, an effective shear-induced horizontal buoyancy force~\cite{rajamanickam2025shear}. Although this force exists even in the Boussinesq approximation, it has been overlooked in the past. Unlike classical vertical buoyancy, which is proportional to the density itself, this emergent force depends fundamentally upon the density gradient.

The purpose of this letter is to demonstrate that this shear-induced force can be formally reinterpreted as the divergence of a Korteweg (1901) stress tensor~\cite{korteweg1901}. By establishing this link, we show that quadratic gradient stresses, traditionally attributed to molecular-scale cohesive potentials and transports, can emerge purely from sub-scale transport processes where the flow is kinematically coupled to the transported gradient. We emphasise that while this derivation is restricted to the narrow-channel, low-order asymptotic regime of~\cite{rajamanickam2025shear}, the mechanism identified here is fundamentally universal. This self-coupling template is expected to emerge in a wide range of multiphysics systems where an internal transport-driven shear flow is kinematically coupled to a scalar gradient. Examples include Marangoni flows driven by local surfactant or temperature variations, geodynamics in porous media, and bio-fluid systems involving chemotaxis or the collective swimming of bacteria in concentration gradients. Consequently, this representation provides a unified mechanical framework for analysing emergent capillarity-like effects across diverse, macroscopically driven transport phenomena.

\section{The depth-averaged two-dimensional description}

Consider a fluid layer of thickness $2h$ confined between two parallel insulating or impermeable planes at $z^*=\pm h$. Gravity acts vertically downwards in the negative $z^*$-direction. The state of the fluid is determined by a scalar field $\theta^*$, representing either temperature or solute concentration, which dictates the density $\rho^*$, through a prescribed equation of state. We assume that the characteristic horizontal length scale $l$ of the scalar field is large compared to the gap thickness $(h\ll l)$,  such that the parameter $\ep =h/l$ is small. The evolution of the scalar field occurs on the diffusive time scale $l^2/D_c^*$, where $D_c^*$ is a reference   diffusion coefficient.

Following the derivation in~\cite{rajamanickam2025shear} with a slight reformulation, the depth-averaged problem is described by the non-dimensional variables and parameters,
\begin{align}
    &t = \frac{t^*}{l^2/\nu_c^*}, \quad \mb x\equiv (x,y) = \frac{1}{l}(x^*,y^*), \quad z=\frac{z^*}{h},  \\ &\widetilde{\mb v} = \frac{\widetilde{\mb v}^*}{\nu_c^*/l}, \quad p = \frac{p^*h^2}{\rho_c^*\nu_c^{*2}},  \quad \theta=\frac{\theta^*}{\theta_c^*}, \quad \rho = \frac{\rho^*}{\rho_c^*}, \\ 
    &\mu = \frac{\rho^* \nu^*}{\rho_c^*\nu_c^*}, \quad \lambda=\frac{\rho^* D^*}{\rho_c^* D_c^*},\quad Gr = \frac{h}{l}\frac{gh^3}{\nu_c^{*2}},\quad \Pra = \frac{\nu_c^*}{D_c^*},
\end{align}
where $Gr$ is the Grashof number and $\Pra$ can be interpreted as the Prandtl number (or Schmidt number). The three-dimensional position and velocity vectors are denoted by $(\mb x,z)$  and $\widetilde{\mb v}=(\mb v,w)$, respectively.   The solution is sought using the perturbation series,
\begin{align}
    \mb v= \ep^{-1}\mb v_0 + \mb v_1, &\quad p = \ep^{-1} p_0 + p_1 +\cdots, \\ w = w_0 +\cdots,&\quad 
    \theta= \theta_0 + \ep \theta_1+\cdots,
\end{align}
in which $\theta_0=\theta_0(\mb x,t)$. The density and transport coefficients are assumed to be of the form $\rho=\rho(\theta)$, $\mu=\mu(\theta)$ and $\lambda=\lambda(\theta)$. The leading-order local flow is the Ostroumov flow or Ostroumov--Birikh--Hansen--Rattray flow~\cite{ostroumov1952svobodnaya,birikh1966thermocapillary,hansen1966gravitational},
\begin{align}
    \mu \mb v_0 &= \frac{Gr}{6}\nabla \rho_0 (z-z^3), \quad \nabla p_0 = Gr \nabla \rho_0 z, \\ w_0 &= \frac{Gr}{24}\nabla\cdot\left(\frac{\rho_0}{\mu_0}\nabla\rho_0\right)(z^2-1)^2.
\end{align}
The fields $(\mb v_0,p_0)$ are antisymmetric about the midplane $z=0$, and vanish upon depth-averaging, $\langle \mb v_0 \rangle = \langle p_0\rangle =0$, where $\langle \cdot \rangle = \tfrac{1}{2}\int_{-1}^{+1}(\cdot) dz$. However, it is precisely this internal shear that generates the dispersion and the buoyancy forces. Furthermore, the Ostroumov flow is simply enslaved to the instantaneous value of local density gradients and viscosity.

The depth-averaged flow field $(\mb u,P)$, which varies on length scales much greater than the channel width, are defined by 
\begin{equation}
    \rho_0 \mb u = \langle \rho_0 \mb v_1 + \rho_1 \mb v_0 \rangle, \qquad 3P=\langle p_1\rangle .
\end{equation}
It is this field that is dynamically coupled to the transport of $\theta_0$, $\rho_0$, $\mu_0$ and $\lambda_0$, which in turn dictates the local internal Ostroumov flow. The large-scale evolution of the depth-averaged system is given by (we drop the subscript ``0" for simplicity)~\cite{rajamanickam2025shear}
\begin{align}
\pfr{\rho}{t} + \nabla \cdot (\rho\mb u) &= 0, \label{cont1}\\
\mu \mb u = -\nabla P &+ \mathbfcal F_\eff, \label{DarcyA} \\
       \rho\pfr{\theta}{t} + \rho \mb u\cdot \nabla \theta   &= \nabla\cdot(\rho \mathbfcal D_\eff \cdot\nabla\theta), \label{theta}\\  
      \rho =\rho(\theta), \quad \mu &= \mu(\theta) \quad \lambda=\lambda(\theta). \label{eqn}
\end{align}
The effective diffusion coefficient $\mathbfcal D_\eff$ and the effective horizontal buoyancy force $\mathbfcal F_\eff$ are defined by
\begin{align}
\rho\mathbfcal D_\eff &=\frac{\lambda}{\Pra}\left(\mathbf I +  \frac{\gamma \Ray^2\rho^2}{\mu^2\lambda^2}\nabla\rho\otimes\nabla\rho\right), \label{Deff1} \\
 \mathbfcal F_\eff &=\gamma  Gr^2\left\{\Pra\,\rho\nabla\left[\frac{(\nabla\rho)^2}{\mu\lambda}\right] +  \frac{3\Pra\rho\nabla\rho}{2\mu^2\lambda} \nabla\rho\cdot\nabla\mu  \right. \nonumber \\ & \quad \left. - \frac{\rho(\nabla\rho\cdot\nabla)}{\mu}\left(\frac{\nabla\rho}{\mu}\right)    - \frac{3\nabla\rho}{2\mu}\nabla\cdot\left(\frac{\rho}{\mu}\nabla\rho\right)\right\},\label{effF1}
\end{align}
where $\Ray=Gr\,\Pra$ is the Rayleigh number and $\gamma=2/2835$ is a numerical factor~\cite{young1991shear}, which emerges as the coefficient when depth-averaging the cross-stream convective-flux term, i.e., $\langle \mb v_0 \theta_1\rangle$. While the Rayleigh number clearly characterises the shear-induced dispersion, the force is better understood in terms of the Grashof number. Equations~\eqref{cont1}-\eqref{effF1}, with suitable initial and boundary conditions, determine the unknowns $\{\mb u,P,\theta,\rho,\mu,\lambda\}$, as functions of $\mb x$ and $t$. 


Evidently, the effective buoyancy force $\mathbfcal F_\eff$ renders the depth-averaged flow field rotational. The origin of the four terms in \eqref{effF1} can be traced to distinct physical mechanisms: the first term represents the tilting and distortion of the hydrostatic balance by the internal flow, while the second term accounts for a similar distortion of the vertical viscous stresses. The remaining two terms, which are independent of the Prandtl number, arise from the depth-averaging of the convective acceleration--the third term stems from the non-linear horizontal momentum flux $\mb v_0\cdot\nabla\mb v_0$ and the fourth is due to the vertical convection of the transverse velocity component $w_0\pfi{\mb v_0}{z}$. While equation~\eqref{effF1} was derived purely from the averaging of the Navier-Stokes equations, its mathematical structure, specifically its dependence on the square and the second-order gradients of density, suggests a deeper physical connection. In the following section, we show that this buoyancy-induced force can be reinterpreted as a divergence of an effective stress tensor, fundamentally linking these confined hydrodynamics to the classical theory of Korteweg stresses.

\section{The effective buoyancy stress as a Korteweg stress}

It is a circumstance of considerable interest that the effective shear-induced force, as defined by equation~\eqref{effF1}, admits a representation in a form formally identical to the classical Korteweg stresses. Although the physical origin of this force lies in the differential hydrostatic balance within the narrow gap of the cell, its mathematical structure reflects the capillary-like stresses employed to model the mechanical effects of density gradients in miscible fluids. The Korteweg stress tensor, $\mb T$, which contributes to the equilibrium portion of the Cauchy stress tensor as  $-P\mb I + \mb T$, is customarily written as~\cite{dunn1986thermomechanics,brenner2014conduction}
\begin{equation}
    \mb T = (\alpha_1 \nabla^2 \rho  + \alpha_2 |\nabla \rho|^2 ) \mb I +\alpha_3 \nabla\rho\otimes\nabla\rho + \alpha_4 \nabla \nabla \rho. \label{kort}
\end{equation}

To establish the correspondence between our derived force and this general stress, we consider the limit of small variations in density and transport properties,
\begin{equation}
    \rho = 1 + \alpha_\rho(\theta-1), \quad \mu = 1 + \alpha_\mu(\theta-1), \quad \lambda = 1+ \alpha_\lambda(\theta-1), \label{eqnstate}
\end{equation}
where the coefficients $\alpha_\rho$, $\alpha_\mu$ and $\alpha_\lambda$ are of the same order of smallness. Neglecting the $O(\alpha_\rho^3)$ terms, the force simplifies to
\begin{equation}
    \frac{\mathbfcal F_\eff}{\gamma Gr^2} =\Pra\,\nabla|\nabla\rho|^2 - (\nabla\rho\cdot\nabla)\nabla\rho    - \frac{3}{2}\nabla\rho\nabla^2\rho 
\end{equation}
which now accounts only for the shear-induced distortion of the hydrostatic balance and the inertial effects of the internal flow. By making use of usual vector identities, we then write $\mathbfcal F_\eff = \nabla \cdot \mb T_\eff$, where
\begin{equation}
   \frac{\mb T_\eff }{\gamma Gr^2}=\left(\Pra+\frac{1}{4}\right) |\nabla \rho|^2  \mb I -\frac{3}{2} \nabla\rho\otimes\nabla\rho. \label{us}
\end{equation}
Comparing this result with the general form in Eq.~\eqref{kort}, we identify the stress coefficients,
\begin{equation}
    \alpha_1=\alpha_4=0, \quad \alpha_2 =  \gamma Gr^2\left(\Pra+\frac{1}{4}\right), \quad \alpha_3 = -\frac{3}{2}\gamma Gr^2.
\end{equation}
The vanishing of $\alpha_1$ and $\alpha_4$ simply implies that the shear-induced force do not have any linear dependence on $\rho$, but only quadratic and higher-order dependences. 

To clarify the physics, we decompose the tensor into isotropic and anisotropic (deviatoric) components,
\begin{equation}
   \frac{\mb T_\eff }{\gamma Gr^2}=\left(Pr-\frac{1}{2}\right) |\nabla \rho|^2  \mb I -\frac{3}{2} \left(\nabla\rho\otimes\nabla\rho-\tfrac{1}{2}|\nabla \rho|^2 \mb I\right). \label{us}
\end{equation}
The isotropic contribution manifests as an \textit{effective fluid pressure}, i.e., $P_\eff\equiv P - \gamma Gr^2 (\Pra-\tfrac{1}{2})|\nabla \rho|^2$. The second term changes sign at a critical Prandtl number and increases the effective fluid pressure for $\Pra<1/2$ and decreases for $\Pra>1/2$. This sign change marks the threshold where the pressure modification transitions from being driven by internal fluid inertia to hydrostatic tilting; the low-$Pr$ regime is typically restricted to liquid metals, such as mercury or liquid gallium, whereas conventional gases and ordinary liquids reside safely in the  $Pr>1/2$ domain.

The anisotropic part, which is a $2\times 2$, symmetric traceless tensor, describes a state of pure shear and has eigenvalues $\pm \tfrac{3}{4}\gamma Gr^2|\nabla\rho|^2$. This tensor exerts a compression along the direction of $\nabla\rho$ and an equal tension in the orthogonal direction. Within the framework of Darcy’s law, where the velocity field is directly proportional to the divergence of the stress, this localized compression effectively ``pushes" the fluid away from regions of high density gradients. Consequently, the net dynamical effect is to dissipate the gradient, acting as a stabilizing, smoothing mechanism. Notably, this anisotropic contribution becomes insignificant at large Prandtl number, $\Pra \to \infty$. In this limit, momentum diffusion is instantaneous relative to mass diffusion, dampening the directional shear induced by buoyancy variations and rendering the effective stress increasingly isotropic.

To further illustrate the mechanical consequences of this representation, consider a circular interface of radius $r=R$ separating two fluid regions of different densities at macroscopic equilibrium $(\mb u=0)$. While a strict mechanical equilibrium is not possible, it is perfectly valid to ignore the $\pfi{\rho}{t}$ in the continuity equation within our low-order asymptotic regime (\(\alpha \ll 1\)) to isolate the primary pressure scaling. Assuming a radial density gradient $\nabla\rho=\rho'(r)\mb e_r$, across a thin transition layer $\sigma \ll R$, the resulting pressure jump $\Delta P=P_{out}-P_{in}$  is obtained by integrating the radial force balance $\pfi{P}{r}=F_{\eff,r}$. The isotropic contribution, being a pure gradient of the squared density gradient, integrates to zero, $\Delta P_{iso}=0$.  This term merely shifts the local pressure within the interface thickness. In contrast, the anisotropic term interacts with the interface curvature to produce a non-vanishing jump
\begin{align}
   \Delta P_{aniso}=-\frac{3\gamma Gr^2}{2R}\int_{R-\sigma}^{R+\sigma} (\rho')^2dr<0.
\end{align}
Crucially, this implies that the anisotropic part of the stress is the primary carrier of the surface tension analogue in this system. Furthermore, because this jump depends on $\rho'^2$, the resulting capillary pressure is strictly independent of the sign of the density contrast; the interface under tension always acts to constrict the interior, whether it be a heavy droplet in a ambient fluid or a light bubble in a heavy fluid. Since this effective stress is born of a miscible system, any such droplet is inherently transient. The Erdogan--Chatwin equation, which governs the transient evolution of $\theta$, shows that an initial $\theta$-jump undergoes an accelerated initial smearing-out where the interface width $\sigma$ grows as $t^{1/4}$ due to the dominance of the shear-induced nonlinear dispersion~\cite{smith1982similarity,grundy1982small}. Given that our pressure jump scales inversely with the interface width, $\Delta P \propto 1/\sigma$, we can infer that the initial transient jump follows a power-law decay
\begin{equation}
\Delta P(t) \sim - \frac{3\gamma Gr^2}{2R} (\Delta \rho)^2 t^{-1/4}.
\end{equation}
This highlights the dual role of the emergent stress: the internal Ostroumov flow that generates the \textit{kinetic surface tension} also serves as the mechanism for gradient relaxation. Consequently, the macroscopic pressure jump vanishes significantly faster than it would under purely molecular diffusion, where $\Delta P \sim t^{-1/2}$~\cite{joseph1996non}. The latter scaling would be realised in the final stages of mixing, when molecular diffusion becomes dominant. For a flat interface ($R\to \infty$), the integrated force across the transition layer vanishes identically, leading to a zero pressure jump $\Delta P=0$. In this planar configuration, the force is perfectly balanced by a local pressure variation within the interface; the force density undergoes a sign change across the inflection point of the density profile.

\subsection{Remark on the hydrodynamic origin of the effective Korteweg stresses}

A fundamental implication of the above correspondence is the insight it provides into the physical nature of gradient-dependent stresses. While the classical Korteweg theory (1901) attributes these stresses to molecular-scale cohesive potentials, our derivation demonstrates that an identical stress tensor emerges from the depth-averaged momentum flux of internal shear. 

This suggests a broader physical conjecture: that the quadratic portion of the Korteweg-type stresses may be a characteristic manifestation of sub-scale transport processes whenever the agent of transport is itself driven by the gradient of the transported field. Specifically, the Ostroumov flow is \textit{kinematically coupled} to the local density gradient. This self-coupling—where the gradient generates the flow that in turn generates the momentum flux—is the essential mechanism that recovers the quadratic gradient terms, $\nabla\rho\otimes \nabla \rho$ and $|\nabla\rho|^2\mb I$. This enslavement means that as the density gradient rotates or evolves, the internal flow and its resulting momentum flux, must rotate and change with it. This allows the depth-averaged stress tensor to ``see" the full geometry of the gradient, leading to the emergence of the full Korteweg structure. 
In this view, these specific "capillarity" coefficients such as $\alpha_2$ and $\alpha_3$, are not mere material constants, but are dynamic properties determined by the ratio of transport rates (such as the Prandtl number) governing the internal relaxation of the interface. 

It is instructive to recall Brenner's bivelocity model~\cite{brenner2012fluid,Brenner2012BeyondN,brenner2014conduction}, which distinguishes between mass and volume velocities to explain the origin of Korteweg stresses. In Brenner’s framework, these stresses are expressed through the gradient and divergence of the volume flux. However, that model remains dependent on a constitutive assumption for the volume flux, linking it to the density gradient, while the resulting stress coefficients remain undetermined material properties.
In our system, a similar \textit{two-velocity} problem naturally arises: the depth-averaged mass velocity $\mb u$ and the internal Ostroumov volume velocity $\mb v_0$. Crucially, unlike the bivelocity model, our constitutive relations and stress coefficients are rigorously derived without phenomenological assumptions, emerging directly from the Navier-Stokes equations in the narrow-channel asymptotic limit. Furthermore, in our derivation, the Laplacian and Hessian contributions to the stress tensor vanish precisely because the underlying internal volume flux is divergence-free, providing a closed-form physical basis for why only the quadratic terms remain.

\subsection{General stress-tensor structure}

So far, we assume small variations in density and transport coefficients to keep the discussions simpler and to make a direct connection to the Korteweg structure~\eqref{kort}. The general stress tensor pertaining to the full force~\eqref{effF1} emerges as $\mb T_\eff=\gamma Gr^2(\mb T_{\text{iso}}+\mb T_{\text{aniso}})$, where
\begin{align}
    &\mb T_{\text{iso}} = \!\left[ \frac{\rho}{\mu}\!\left(\frac{Pr}{\lambda}\!-\!\frac{1}{2}\right)\!|\nabla\rho|^2 \!+ \!\! \int \! \frac{Pr}{\mu^2\lambda}|\nabla \rho|^2\!\left(\frac{3\rho}{2}\mathrm{d}\mu-\mu\mathrm{d}\rho\right) \right]\!\mb I,  \\
    &\mb T_{\text{aniso}} =-\frac{3\rho}{2\mu^2}\left(\nabla\rho\otimes\nabla\rho-\tfrac{1}{2}|\nabla \rho|^2 \mb I\right). \label{ansioT}
\end{align}
Unlike previous cases where the stress coefficients are constants, the full non-linear tensor reveals strongly state-dependent coefficients and an inherited integral closure over the transport properties. This formulation can be interpreted as an extended Korteweg structure that incorporates new non-local contributions not present in the classical Korteweg framework. The non-local term in $\mb T_{\text{iso}}$ represents a path-independent line integral in state space (or property space) since $\rho$, $\mu$, and $\lambda$ are uniquely parameterised by $\theta$. The new term being cubic in differentials $|\nabla \rho|^2d\rho$ or $|\nabla\rho|^2d\mu$ did not arise in our earlier discussions for $\alpha_\rho \ll 1$. More importantly, the kinetic surface tension which originates from $\mb T_{\text{aniso}}$ now indicates a capillary coefficient that depends on $\rho$ and $\mu$.

\subsection{The effective shear-induced force for classical Taylor dispersion}

To further distinguish the unique nature of the self-coupled Ostroumov flow, we contrast it with the effective force arising in classical Taylor dispersion. While deriving this force for varying viscosity remains an open problem, the constant-viscosity case $\mu=1$ (and $\lambda=1$ for simplicity), provides a clear counter-example. Following the frameworks established in~\cite{rajamanickam2022effects,rajamanickam2023thick,rajamanickam2024effect}, we consider a Poiseuille flow in a Hele-Shaw cell with mean speed $U$ directed along the unit vector  $\mb e$. In a frame moving with the mean flow, the depth-averaging of the non-Boussinesq momentum equations yields the effective force~\cite{rajamanickam2024effect}
\begin{align}
    \frac{\mathbfcal F_\eff}{\gamma Re^2} =- \left(\Pra+3\right) (\mb e\cdot \nabla\rho) \mb e. 
\end{align}
This implies
\begin{equation}
    \frac{\mb T_\eff }{\gamma Re^2}= - \left(\Pra+3\right) \rho\, \mb e\otimes \mb e,
\end{equation}
where $\gamma=2/105$ and $Re=Uh/\nu_c^*$ is the Reynolds number; when $\lambda$ is allowed to vary, the force takes the form $\mathbfcal F_\eff=-\gamma Re^2(\tfrac{\Pra}{\lambda}+3)(\mb e\cdot \nabla\rho) \mb e$. Alternatively, we may also consider a Couette flow in a Hele-Shaw cell with zero mean flow, say, a local description of a Taylor-Couette configuration for small annular gaps. For this scenario, we can allow the transport coefficients to vary. If $U$ represents the velocity of the wall motion, then we find
\begin{align}
    &\frac{\mathbfcal F_\eff}{\gamma Re^2} =- \left(\frac{\mu\Pra}{\lambda}+1\right) (\mb e\cdot \nabla\rho) \mb e - \frac{\rho}{\lambda}(\mb e\cdot \nabla\mu) \mb e,
\end{align}
where $\gamma=2/15$. This force can be expressed as the divergence of the tensor
\begin{equation}
    \frac{\mb T_\eff }{\gamma Re^2}= - \left[\rho+\int \frac{Pr}{\lambda}(\mu d\rho + \rho d\mu)\right]  \mb e\otimes \mb e,
\end{equation}
which simplifies to $\mb T_\eff=-\gamma Re^2 (Pr+1)\mb e\otimes \mb e$ when $\mu=\lambda=1$.

Unlike the Korteweg stresses derived in Section 3.1, this Taylor-dispersion stress tensor is strictly uniaxial, acting only along the mean-flow direction $\mb e$.  Because the Poiseuille profile is an externally driven background field independent of the scalar gradient, it lacks the self-coupling necessary to produce the full $(\nabla\rho\otimes\nabla\rho)$ Korteweg structure.  This comparison reinforces our central thesis that capillarity-like Korteweg stresses are a specific manifestation of systems where the internal transport--shear in our case--is kinematically coupled to the gradient of the transported field.

\section{Concluding remarks}

In this letter, we have demonstrated that the shear-induced buoyancy forces emerging in non-Boussinesq, depth-averaged flows admit a representation formally equivalent to a Korteweg stress tensor. This correspondence provides a rigorous hydrodynamic origin for quadratic density-gradient stresses, showing they can arise purely from the self-coupling between a scalar gradient and the internal (Ostroumov) flow it drives.

It is noteworthy that the emergent stress tensor derived here cannot be recovered from a standard Cahn--Hilliard or Ginzburg--Landau energy functional. For instance, in typical diffuse-interface fluid models based on the energy density of the form $f = f_0(\rho) + \frac{\chi}{2}|\nabla \rho|^2$, the variational framework strictly dictates the structural pairing of various terms leading to $\mb T = \chi \nabla\rho\otimes\nabla\rho - \chi \rho \nabla^2\rho \mb I$, locking the stress coefficients to a single capillary constant $\chi$~\cite{giovangigli2020kinetic}. More generally, the stress coefficients are strictly constrained by the requirement of thermodynamic consistency, most notably through the relations identified by Dunn and Serrin~\cite{dunn1986thermomechanics}, leading to $\mb T = \chi(\rho) \nabla\rho\otimes\nabla\rho -  \rho\chi(\rho) \nabla^2\rho \mb I-\rho \frac{\mathrm{d}\chi}{\mathrm{d}\rho}|\nabla \rho|^2\mb I$, which are not satisfied by the hydrodynamic coefficients identified here. This structural mismatch implies that the buoyancy-induced stress is inherently non-variational; it represents a purely kinetic phenomenon born of momentum flux rather than the minimisation of a free-energy potential. Additionally, the specific algebraic pairing of the classical coefficients ensures that a static equilibrium ($-\nabla P + \nabla \cdot \mb T = \mathbf{0}$) of an interface exists by guaranteeing that $\nabla \cdot \mb T = \rho \nabla \Psi$, so that the vector equilibrium equation reduces to $-\nabla \mu_c + \nabla \Psi = \mathbf{0}$ with $\nabla \mu_c = \tfrac{1}{\rho}\nabla P$, i.e., a problem for a single scalar function $\rho$. In our case, this structural mismatch implies a non-vanishing curl for the effective force, meaning a general mechanical equilibrium is impossible unless a Boussinesq limit ($\alpha \ll 1$) is considered under highly symmetric configurations (see~\cite{serrin1983form}); otherwise, the interface must continuously evolve via hydrodynamic convection. Consequently, our results identify a class of `kinetic' Korteweg stresses that sit outside the classical potential-based framework, providing a new perspective on the mechanics of miscible interfaces. It is worth noting that standard derivations of the Korteweg stress tensor for dense gases via higher-order terms in the Chapman--Enskog expansion exist in the literature~\cite{rocard1952thermodynamique,giovangigli2020kinetic}. The emphasis here is the purely continuum hydrodynamic origin of the stress, which is also anticipated to exist in other transport configurations mentioned in the Introduction, areas which require future attention. For instance, the active stress tensor, $\mb T_{\text{active}}=-\zeta \mb Q$ 
(with $\mb Q = s(\mb n\otimes \mb n-\tfrac{1}{2}\mb I)$), in two-dimensional 
active matter bears a close resemblance to the anisotropic part of the Korteweg tensor when activity is associated with localized spatial variations of a single scalar field, say $\rho$. Making the correspondence to our case, we may write
\begin{equation}
    \zeta = \frac{3\gamma Gr^2\rho}{2\mu^2}, \qquad s = |\nabla \rho|^2, \qquad \mb n = \frac{\nabla \rho}{|\nabla \rho|},
\end{equation}
which indicates that the shear-induced buoyancy stress corresponds to an extensile active-matter-type fluid since $\zeta>0$.

Finally, the transient nature of these stresses reflects the buoyancy-driven transport inherent to the system. Unlike classical surface tension, the pressure jump across a miscible interface in this system undergoes an accelerated power-law decay, $\Delta P \sim t^{-1/4}$, as the same buoyancy-induced shear that generates the stress also drives the rapid smoothing of the density gradient. This specific scaling behaviour is suggested for validation through future microfluidic experiments or high-resolution 3D numerical simulations. Moreover, a complete description of Korteweg-type stresses in compressible or variable-property fluids may be more general than the classical form, notably through a dependence on viscosity gradients, a direction that warrants further theoretical consideration.

\bibliography{apssamp}

\end{document}